# Optimizing single Slater determinant for Electronic Hamiltonian with Lagrange multipliers and Newton-Raphson methods as an alternative to ground state calculations via Hartree-Fock self consistent field

Sandor Kristyan

*Research Centre for Natural Sciences, Hungarian Academy of Sciences,*
*Institute of Materials and Environmental Chemistry*
*Magyar tudósok körútja 2, Budapest H-1117, Hungary,*

Corresponding author: kristyan.sandor@ttk.mta.hu

**Abstract**. Considering the emblematic Hartree-Fock (HF) energy expression with single Slater determinant and the ortho-normal molecular orbits (MO) in it, expressed as a linear combination (LC) of atomic orbits (LCAO) basis set functions, the HF energy expression is in fact a 4$^{th}$ order polynomial of the LCAO coefficients, which is relatively easy to handle. The energy optimization via the Variation Principle can be made with a Lagrange multiplier method to keep the ortho-normal property and the Newton-Raphson (NR) method to find the function minimum. It is an alternative to the widely applied HF self consistent field (HF-SCF) method which is based on unitary transformations and eigensolver during the SCF, and seems to have more convenient convergence property. This method is demonstrated for closed shell (even number of electrons and all MO are occupied with both, α and β spin electrons) and restricted (all MOs have single individual spatial orbital), but the extension of the method to open shell and/or unrestricted cases is straightforward.

**Keywords.** Alternative method for Hartree-Fock self consistent field,
Lagrange multipliers method to optimize single Slater determinant

## INTRODUCTION

The widely applied standard HF-SCF approximate solution [1-3] of the non-relativistic electronic Schrödinger equation for an N-electron molecular system (containing A=1,…,M atoms with $Z_A$ nuclear charges at positions $\mathbf{R}_A$ [4-6]) approximates the total ground state electronic energy with single Slater determinant ($\Psi_0 \approx S_0$), and the energy expression for closed shells with restricted MOs in $S_0$ is

$$E_{0,total}^{HF-SCF} = 2 \Sigma_{i=1}^{N/2} (i|h|i) + a\Sigma_{i=1}^{N/2}\Sigma_{j=1}^{N/2} (2(ii|jj)-(ij|ji)) + V_{nn} \quad (1)$$

(see p.68, equation 3.127 on p.134 in ref.[1]. For example, using $i(1) \equiv i(\mathbf{r}_1)$ and $j(1) \equiv j(\mathbf{r}_1)$ spatial functions for pairwise occupation: if N/2=1 ⇒ $S_0= |\alpha_1 i(1), \beta_2 i(2)>= (\alpha_1\beta_2-\alpha_2\beta_1)i(1)i(2)$, if N/2=2 ⇒ $S_0= |\alpha_1 i(1), \beta_2 i(2), \alpha_3 j(3), \beta_4 j(4)>$, generally $i(1) \equiv i(\mathbf{r}_1)$ the i$^{th}$ MO, with i=1,2,…N/2 as a function of position vector $\mathbf{r}_1=(x_1,y_1,z_1)$ of electron 1, as commonly abbreviated in computational quantum chemistry; unusual, (but useful in sums) is the function notation i(1) instead of f($\mathbf{r}_1$). The h($\mathbf{r}_1$)= -(1/2)$\nabla_1^2$ - $\Sigma_{A=1…M} Z_A/R_{A1}$ with sum and $R_{A1} \equiv |\mathbf{r}_1-\mathbf{R}_A|$ is the kinetic energy (T) plus Coulomb nuclear-electron attraction ($V_{ne}$) operator, the $\Sigma_i\Sigma_j$ term approximates the Coulomb electron-electron repulsion energy ($V_{ee} \equiv$ (N(N-1)/2)$\int\Psi_0^*\Psi_0 r_{12}^{-1/2}$ $\Pi_{n=1}^N ds_n d\mathbf{r}_n$, for which a particular closed shell restricted MOs case is written out), and $V_{nn}= \Sigma_{A=1…M} \Sigma_{B=A+1…M} Z_A Z_B/R_{AB}$ is the Coulomb nuclear-nuclear repulsion energy. The parameter 'a' in Eq.1 is the "coupling strength parameter": a=1 has a physical meaning, while a=0 means that there is no electron-electron interaction useful for mathematical purposes [7-8]. Generally, $(i|h|j) \equiv \int i^*(1)hj(1)d\mathbf{r}_1$ and $(ij|kl) \equiv \int i^*(1)j(1)r_{12}^{-1} k^*(2)l(2) d\mathbf{r}_1 d\mathbf{r}_2$, which simplifies a bit for real functions ($i(1)^*=i(1)$) what one has for MOs, as well as the integration by parts provides for u=x,y,z: -$\int i(1)(\partial^2/\partial u_1^2)j(1)d\mathbf{r}_1= \int(\partial i(1)/\partial u_1)(\partial j(1)/\partial u_1)d\mathbf{r}_1$, because the MOs are well behaved (square integrable and tend to zero as $|\mathbf{r}_1|\to\infty$). The ground state one-electron density ($\rho_0$, with $s_n= \alpha$ or β spin state of electron n=1…N) and its closed shell approximation is

$$\rho_0(\mathbf{r}_1) \equiv N\int\Psi_0^*\Psi_0 \, ds_1\Pi_{n=2}^N ds_n d\mathbf{r}_n \quad \text{and} \quad \rho_0^{HF-SCF}(\mathbf{r}_1) \equiv 2\Sigma_{i=1}^{N/2} i(1)^2 \quad \text{with} \quad i(1)= \Sigma_{k=1…K} c_{ik} G_k(1). \quad (2)$$

For open shell case (N is odd or ∃ not-closed MO), the energy expression is similar to Eq.1 and can be discussed analogously; $\rho(1)$ is an approximate for $\rho_0$, e.g. $\rho(1):= \rho_0^{HF-SCF}(\mathbf{r}_1)$ or else (e.g. Appendix.4).

The LCAO approximation for the i=1,2…N/2 MOs is given in Eq.2, where the set $\{c_{ik}\}$ contains the LCAO coefficients to the basis set $\{G_k\}_{k=1,…K}$, and $K \geq N/2$ must be. With MO energy, increasing upward, picture the known arrangement of the rectangular matrix of LCAO coefficients as

| (N/2)$^{th}$ MO | | $c_{N/2\ 1}$=1 | * | * | * | * | * | * | * | $c_{N/2\ K}$ |
|---|---|---|---|---|---|---|---|---|---|---|
| | | * | *=1 | * | * | * | * | * | * | * |
| i$^{th}$ MO | | * | * | *=1 | * | * | * | * | * | * |
| | | * | * | * | *=1 | * | * | * | * | * |
| 2$^{nd}$ MO | | $c_{21}$ | * | * | * | *=1 | * | * | * | $c_{2K}$ |
| 1$^{st}$ MO (ground) | | $c_{11}$ | * | * | * | * | *=1 | * | * | $c_{1K}$ |

The number of elements in set $\{c_{ik}\}$ is NK/2. Generally, the atom centered $G_k(1) \equiv (x_1-R_{Ax})^{nx} (y_1-R_{Ay})^{ny} (z_1-R_{Az})^{nz} \exp(-b|\mathbf{r}_1-\mathbf{R}_A|^m)$ Gaussian type orbits (GTO, m=2) for which analytical integration is available, or the more effective, but not easy to handle Slater type orbits (STO, m=1) are chosen. (Contracted GTO are also used as $G_k$, as more powerful basis set, etc. not detailed here, also the concepts like minimal basis, STO-3G (STO approximated with three GTO), etc..) For the demonstration we use all the atomic STO basis functions (p.88 and pp.92-94 in ref.[9]) up to (n,l)= (3,+2) quantum numbers for all individual atoms up to $Z_A$=18 (Ar), (non-relativistic cases), and numerical integration for all cross products (see below).

For initial LCAO values in the iteration for Eqs.4-5 below, one can use the sophisticated "Harris approximation" used in HF-SCF practice, or simply, mainly because stationary (equilibrium geometry) close to neutral molecules are in focus, our choice is as follow: e.g. for a quasi-neutral C atom ($N_A$=6 from N= $\Sigma_{A=1}^M N_A$) in molecule the $c_{ik}$= 1.0 for occupied 1s, 2s, the $c_{ik}$= 1.0 (or the finer 2/3) for partially occupied $2p_x$, $2p_y$, $2p_z$ and $c_{ik}$= 0.0 for higher, unoccupied excited AOs (3s, $3p_x$, etc.), based on the configuration of ground state atomic C ($1s^2 2s^2 2p_x^1 2p_y^1$). Based on some concepts of chemical bonds (core electrons, etc.), this means that in a proper matrix arrangement (Appendix.1) the $c_{ik}$= $\delta(i,k)$= 1 if i=k and 0 if i≠k, the Kronecker delta (see the values 1 as initial values for the $\{c_{ik}\}$ matrix above), and some $c_{ii}$=1 are zeroed out as explained in the case of a neutral C atom; a chemically plausible choice for initial $\{c_{ik}\}$. In case of a well chosen "minimal basis", the $\{G_k\}$ can be partitioned to atoms it centered as $U_{A=1}^M\{G_k(\text{centered on } A)\}$, wherein all subset contains ortho-normalized AOs (that is, solutions of the Schrödinger equation with M=1, that is, these initial $\{c_{ik}\}$ satisfy all the diagonal and some off-diagonal equations in Eq.5 below), and as seen in the practice, in this hierarchy, the values in minimizing set $\{c_{ik}\}$ are in or not far (Apendix.1) from the interval [-1,1], a convenient property for the iteration below.

**Energy optimization with Lagrange multipliers**

The idea is based on the fact that inserting the LCAO approximation for i(1) in Eq.2 into Eq.1 yields a 4$^{th}$ order multivariable polynomial of $\{c_{ik}\}$ if a=1, 2$^{nd}$ order only if a=0, and the Lagrangian (keeping the N/2 MOs ortho-normal) is

$$L= 2\Sigma_{i=1}^{N/2} \int i(1) h\ i(1) d\mathbf{r}_1 + a\Sigma_{i=1}^{N/2}\Sigma_{j=1}^{N/2} \int (\ 2i(1)^2 j(2)^2 - i(1)j(1)i(2)j(2)\ ) r_{12}^{-1}\ d\mathbf{r}_1 d\mathbf{r}_2 +$$
$$+ \Sigma_{i=1}^{N/2}\Sigma_{j=i}^{N/2} \lambda_{ij}(\int i(1)j(1)d\mathbf{r}_1 - \delta(i,j)) + V_{nn}\ . \quad (3)$$

The $\lambda_{ij}$ are the Lagrange multipliers [10], only an upper diagonal matrix (i.e. j runs in [i, N/2] only), and $\delta(i,j)$ is the Kronecker delta. Picture the multipliers $\{\lambda_{ij}\}$ as an upper triangle, square matrix as

| for 1$^{st}$ vs j$^{th}$ MO | | $\lambda_{11}$ | $\lambda_{12}$ | * | $\lambda_{1\ N/2}$ |
|---|---|---|---|---|---|
| | | - | * | * | * |
| | | - | - | * | * |
| for (N/2)$^{th}$ MO vs. itself | | - | - | - | $\lambda_{N/2\ N/2}$ |

The number of elements in set $\{\lambda_{ij}\}$ is the number of diagonal plus upper diagonal elements in an (N/2)x(N/2) matrix: N(N+2)/8, the proof is elementary. Energy minimization is to find the minimum of L with respect to parameters $\{c_{ik}\}$ and $\{\lambda_{ij}\}$, where i,j=1…N/2 with j≥i runs for MOs and k=1…K runs for basis set functions. The polynomial order of parameters {LCAO coefficients}U{Lagrange multipliers}=

$\{c_{ik}\}\cup\{\lambda_{ij}\}$ of four main terms in Eq.3 is $2^{nd}$, $4^{th}$, $3^{rd}$ and $0^{th}$, the last $V_{nn}$ term is only an additive constant. The minimum of L is obtained by solving the nonlinear system $\{\partial L/\partial c_{ik}=0, \partial L/\partial \lambda_{ij}=0$ for all i,j,k$\}$, e.g. with NR method (p.263 of ref.[11]), for which the second derivatives of L, (the $\partial^2 L/\partial c_{ik}\partial c_{jk'}$, $\partial^2 L/\partial \lambda_{ij}\partial \lambda_{i'j'}$ and $\partial^2 L/\partial c_{ik}\partial \lambda_{i'j}$) are also necessary. The derivation is straightforward, but can be done simpler if the double sums are shifted inside to have terms like $\Sigma_{i=1}^{N/2}i(1)^2\Sigma_{j=1}^{N/2}j(2)^2$ along with $\partial(\Sigma_{i=1}^{N/2}i(1)^2)/\partial c_{ik}= 2i(1)G_k$, etc.:

$$0= g_{ik}\equiv \partial L/\partial c_{ik} = 4\int G_k(1)\, h\, i(1)\, d\mathbf{r}_1 + \int G_k(1)[i(1)\lambda_{ii} + \Sigma_{j=1}^{N/2}\lambda_{ij}j(1)]d\mathbf{r}_1 +$$
$$+2a\Sigma_{j=1}^{N/2}\int\{4G_k(1)i(1)j(2)^2 - [G_k(1)i(2)+G_k(2)i(1)]j(1)j(2)\}r_{12}^{-1}d\mathbf{r}_1d\mathbf{r}_2 , \qquad (4)$$
$$0= h_{ij}\equiv \partial L/\partial \lambda_{ij} = \int i(1)j(1)d\mathbf{r}_1 - \delta(i,j) \qquad (j\geq i) . \qquad (5)$$

Notice the hectic double indexing in $\lambda_{ij}$ in Eq.4: e.g. $\lambda_{21}, \lambda_{22}, \lambda_{23}$ for i=2 and N/2=3, it should be considered as $\lambda_{12}, \lambda_{22}, \lambda_{23}$, since $j\geq i$; i.e. picking from upper diagonal. (Careless doubling the terms with $\lambda_{ji}$ beside $\lambda_{ij}$ is not allowed, because that would cause singular matrix in Eq.13 below. For the Jacobian, $\partial h_{ij}/\partial \lambda_{i'j'}=0$, $\partial h_{ij}/\partial c_{ik}=\int G_k(1)j(1)d\mathbf{r}_1$ and $\partial h_{ij}/\partial c_{jk}=\int G_k(1)i(1)d\mathbf{r}_1$ if j>i, $\partial h_{ii}/\partial c_{ik}=2\int G_k(1)i(1)d\mathbf{r}_1$, $\partial h_{ij}/\partial c_{i'k}=0$ if $i\neq i'\neq j$, and similarly, $\partial^2 L(a=0)/\partial c_{ik}^2= \partial g_{ik}(a=0)/\partial c_{ik}= 2\int G_k[2h+\lambda_{ii}]G_k d\mathbf{r}_1$, etc..) The initial parameters for the set $\{\lambda_{ij}\}$ can be calculated from Eq.4 using the initial parameters chosen for the set $\{c_{ik}\}$ above, because Eq.4 is a simple linear system for $\{\lambda_{ij}\}$, as well as notice that, Eq.4 has more equations than necessary for this step: Pick N(N+2)/8 from NK/2. The grad(L)= $(\partial L/\partial c_{11},..., \partial L/\partial \lambda_{N/2,N/2})$ via Eqs.4-5 shows the opposite direction in which L decreases most quickly, and |grad L| determines how fast the L changes in that direction in $\{c_{ik}\}\cup\{\lambda_{ij}\}$ space.

We minimize Eq.1 first with a=0 via the LCAO coefficients $\{c_{ik}\}$, that is solving Eqs.4-5 for zero $1^{st}$ derivatives with NR method (with the help of Eqs.7-13 below) for $\{c_{ik}\}\cup\{\lambda_{ij}\}$, and use it as initial parameters for the wanted case when a=1 (but doing the same computation procedure). The reason for these two steps is that Eq.3 is only $2^{nd}$ order in $\{c_{ik}\}$ and $3^{rd}$ order in $\{c_{ik}\}\cup\{\lambda_{ij}\}$ if a=0, and the zero $1^{st}$ derivatives (Eqs.4-5) can be found in stable ways, as well as the stationary set $\{c_{ik}\}$ does not change too much [7-8] in a$\in$[0,1] (see Appendix.1). Strictly saying, the "perturbation theory" comprises mathematical methods for finding an approximate solution to a problem, by starting from the exact solution of a related, simpler problem, and in fact this pre-calculation trick with parameter 'a' can also be considered as a perturbation, see Appendices 1-2. A weaker approximation [1-2] than the fully minimized Eq.1 with a=1 can be calculated by LCAO coefficients from Eqs.4-13 at a=0, and using directly in Eq.1 along with a=1. We focus on ground state here, but the MOs at a=0 can be used as a basis for "configuration interactions (CI)" methods for ground- and excited states, wherein the off-diagonal elements in the CI matrix depend on operator $r_{12}^{-1}$ only, for this, one must introduce at least one/two additional virtual $(1+N/2)^{th}$ electrons [7-8].

To solve the system in Eqs.4-5, we need the first derivatives for $g_{ik}$ and $h_{ij}$, which is straightforward. The $\partial h_{ij}/\partial \lambda_{i'j'}= 0$ since L is linear in $\lambda_{ij}$, a lucky situation, but the two groups of functions (g and h) with double indexing (ik and ij) along with $j\geq i$ is hectic, so the transformation of indexing is useful, see below. We see from Eqs.3-5 that the powers build up as

$$L= \Sigma c_{ik}c_{i'k'} + \Sigma c_{ik}c_{i'k'}\lambda_{i''j} + a\Sigma c_{ik}c_{i'k'}c_{i''k''}c_{i'''k'''} \qquad (6)$$
$$g_{ik}= \Sigma c_{ik} + \Sigma c_{ik}\lambda_{i'j} + a\Sigma c_{ik}c_{i'k'}c_{i''k''} \quad \text{and} \quad h_{ij}= \Sigma c_{ik}c_{i'k'} \qquad (7)$$

where the symbolic $\Sigma$ means proper LC with indices run, and the coefficients in these individual LCs can be mixed up from the elementary integrals to be calculated in Eqs.3-5, which are the same as in standard HF-SCF procedure: Two center integrals $\int G_k(1)G_{k'}(1)d\mathbf{r}_1$ from Eq.5 and the two and three center

$$\int G_k(1)\, h\, G_{k'}(1)d\mathbf{r}_1 = -(1/2)\int G_k(1)\nabla_1^2 G_{k'}(1)d\mathbf{r}_1 - \int G_k(1)G_{k'}(1)(\Sigma_{A=1...M} Z_A/R_{A1})d\mathbf{r}_1 \qquad (8)$$

(first and second term in the right, resp.) integrals from Eq.4 along with the four center integrals

$$\int G_k(1)G_{k'}(1)G_{k''}(2)G_{k'''}(2)r_{12}^{-1}d\mathbf{r}_1 d\mathbf{r}_2, \qquad (9)$$

where $G_k$'s are primitive (or contracted) Gaussians, see Appendix.3-4. Eq.7 shows the degree of multi-variable polynomials (g is $3^{rd}$ order, h is $2^{nd}$ order) and the missing powers (g has no squares like $c_{ik}c_{i'k'}$, h has no linear terms like $c_{ik}$ or $\lambda_{ij}$ etc.). In the first step calculation with a=0, the system in Eq.7 is second order (parabolic) only to solve for minimizing L in Eq.3. (In the $1^{st}$ sum in Eq.6 i=i' in $c_{ik}c_{i'k'}$ by Eq.3, etc..)

The set $\{c_{ik}\}\cup\{\lambda_{ij}\}$ of variables in the multi-variable polynomials in Eqs.4-5 is not easy to handle because the two indices, again, i=1…N/2, j=i…N/2 for (i) and (j) MOs where the number of electrons N is even, as well as k=1…K for basis set functions $\{G_k\}$. It is useful to transfer them to one index variables: Let $f_n= g_{ik}$ for functions and $x_n= c_{ik}$ for LCAO coefficients, where

$$n=(i-1)K + k \qquad (10)$$

ran as n=1, 2, …NK/2. For the other functions $h_{ij}$ and Lagrange multipliers $\lambda_{11},…, \lambda_{1,N/2}, \lambda_{22}…, \lambda_{2,N/2}, \lambda_{33}…, \lambda_{3,N/2}, …, \lambda_{N/2,N/2}$ the n in $f_n= h_{ij}$ and $x_n= \lambda_{ij}$ continues as

$$n= NK/2+1, NK/2+2, NK/2+3……P\equiv NK/2+N(N+2)/8= (N/2)(K+N/4+1/2), \qquad (11)$$

respectively. ($N(N+2)/8$ is the cardinality of the set $\{\lambda_{ij}\}$.) Eqs.10-11 also show that the number of variables (dimensionality) in the 4$^{th}$ order polynomial $L(c_{11},…, \lambda_{N/2,N/2})=L(x_1,…,x_P)$ in Eq.3 is P in Eq.11 with even N. The left hand side of the equation system in Eqs.4-5 is a zero column vector, but the i=j=1,…N/2 cases of Eq.5 rearrange it as $1= \int i(1)^2 d\mathbf{r}_1$, providing non-singular system. Finally, the system in Eq.7 transfers (with the analogue meaning for symbolic Σ) to

$$0= f_n= \Sigma x_{n1} + \Sigma x_{n1} x_{n2} + a\Sigma x_{n1} x_{n2} x_{n3} \qquad (12)$$

where $f_n= f_n(x_1,….x_P)$ with n, $n_1$, $n_2$, $n_3$=1…P, with P in Eq.11. This transfer of indices for Eq.7, which can even start at Eq.6 as $L= \Sigma x_{n1} x_{n2} + \Sigma x_{n1} x_{n2} x_{n3} + a\Sigma x_{n1} x_{n2} x_{n3} x_{n4}$ with $0= f_n\equiv \partial L/\partial x_n$, avoids the caution j≥i for $\lambda_{ij}$ mentioned after Eq.4. In this way, the polynomial generation in Eq.12, for example, for the term $a_n x_1^2 x_4$ comes from $b_{114}x_1 x_1 x_4+ b_{141}x_1 x_4 x_1+ b_{411}x_4 x_1 x_1$ with $a_n=b_{114}+b_{141}+b_{411}$. This transfer of indices in Eqs.10-11 can be conveniently done with e.g. FORTRAN "do-loops" in the programming.

Let $[x_n(0)]$ be the column vector of the initial values for $x_n$, the function value at this point is the column vector $[f_n(0)]$, then with a PxP iteration matrix $[M(m)]$, the iteration [11] for m=0,1,2… is

$$[x_n(m+1)]= [x_n(m)] + [M(m)][f_n(m)] . \qquad (13)$$

It avoids the eigensolver and unitary transformation what e.g. HF-SCF uses to find the minimizing LCAO coefficients. Eq.13 should be applied for pre-optimization (a=0 in Eqs.4-5, 2$^{nd}$ order polynomial) first, then a re-optimization for a=1 in Eqs.4-5 (4$^{th}$ order). The choices for [M] in Eq.13 are: 1.: Unit matrix, called "iterative method", however, a strong condition in this case requires the initial $[x_n(0)]$ to be very close to the optimum, generally not feasible (may be in the re-optimization step), 2.: The NR as $[M(m)]= -[W(m)]^{-1}$, where $[W]\equiv [\partial f_{n1}/\partial x_{n2}]$ is the Jacobian matrix of Eqs.4-5 (or Hessian of L in Eq.3), 3.: $-1/[\partial f_{n1}/\partial x_{n1}]_m$ diagonal and zero off-diagonal elements in [M(m)], called "diagonal NR", particularly, $\partial^2 L/\partial \lambda_{ij}^2=0$ from Eq.5 does not allow this here, 4.: The fast "gradient (steepest descent)" method converging along -grad(L) if $[M(m)]= -b(m)[W(m)]^T$, where $b= [f_n]^T [q]/([q]^T [q])$ and $[q]\equiv [W][W]^T [f_n]$, (even matrix inversion is not necessary, only transpose, more, since W is Hessian ⇒ W is symmetric ⇒ W=W$^T$), 5.: The "diagonal gradient" neglects the off-diagonal part of W yielding $b=(\Sigma_{n=1}^P (\partial f_n/\partial x_n)^2 f_n^2)/(\Sigma_{n=1}^P (\partial f_n/\partial x_n)^4 f_n^2)$, but with slower convergence, however, the sum in the denominator avoids the problem of e.g. $\partial^2 L/\partial \lambda_{ij}^2=0$. Notice that if P=1, the "diagonal gradient" reduces to $b=(\partial f_1/\partial x_1)^{-2}$ and 1x1 size $[W]=\partial f_1/\partial x_1$, so $-b[W]^T[f_1]= -f_1/(\partial f_1/\partial x_1)$, the 1 dimensional NR. In cases 1, 3 and 5, even the transfer of variables (Eqs.10-11) would not be necessary and the use of PxP matrix [M] in Eq.13 would reduce with the use of Px1 column vector $[f_n]$.

## APPENDIX

**Appendix.1 The initial LCAO coefficients**: For example, in case of LiH molecule, the HF-SCF expands the 2x6 matrix $\{c_{ik}\}$ to 6x6 matrix (virtual orbits) and uses unitary transformation, etc., while the method with Eqs.3-5 uses the 2x6 matrix $\{c_{ik}\}$ with 2x2 upper triangle matrix $\{\lambda_{ij}\}$:

Simple initial LCAO (atomic H config.: $1s^1 \Rightarrow c_{12}:=1.0$, atomic Li config.: $1s^2 2s^1 \Rightarrow c_{21}:=1.0$, N=4):

```
    basis  Li(1s)      H(1s)     Li(2s)    Li(2px) Li(2py) Li(2pz)
2.MO:       1.0         0          0         0       0       0
1.MO:       0           1.0        0         0       0       0
```

Converged LCAO by HF-SCF/STO-3G/a=0 for Eq.1 ($E_{0,total}$= -11.456970 hartree, hypothetic bound state):

```
2.MO:       1.00550     0.00435   -0.02504   0.0     0.0    -0.00159
1.MO:      -0.10689     0.66300    0.30999   0.0     0.0    -0.32185
```

Converged LCAO by HF-SCF/STO-3G/a=1 for Eq.1 ($E_{0,total}$= -7.860313 hartree, real bound state):

```
2.MO:      -0.99129    -0.00349   -0.03290   0.0     0.0    -0.00603
1.MO:       0.16462    -0.55155   -0.45875   0.0     0.0     0.34460
```

Notice that, the two sets of converged $\{|c_{ik}|\}$ are similar (apart from -1 phase factor).

**Appendix.2 The ratios of energy terms**: The term $E_0\equiv E_{0,total}-V_{nn}= T + V_{ne} + V_{ee}$ is called electronic energy in a real case (a=1 in Eq.1), and the ratios in the right side is interesting in relation to the parameter 'a'. With the help of small $H_2$ and larger $C_{10}H_8$ equilibrium geometry molecules we demonstrate that, the magnitude of V&T energy ratios in Eq.3 does not depend strongly on N. The calculation below represents the involvement of $V_{ee}$ interesting for convergence in view of manipulation with parameter 'a' in Eq.4;

notice that, operator 'h' in Eq.3 counts for T+$V_{ne}$, the |$V_{ne}$| > T, $V_{ee}$ > 0 and $V_{ne}$ < 0 always in real systems, as well as the virial ratio -($V_{ne}$+$V_{ee}$+$V_{nn}$)/T=2 has the strict value 2 (a=1 or a≠1), not the listed ones:

**TABLE 1.** Energy ratios

| All energies (except $V_{nn}$) are HF-SCF/STO-3G/a=1 level | $E_{0,total}$ HF-SCF/STO-3G $V_{nn}$ [hartree] | $V_{ee}$ /(T+$V_{ne}$) | T : $V_{ne}$ : $V_{ee}$ ratio normalized to $V_{ne}$:=-1 |
|---|---|---|---|
| H$_2$ (hydrogen, N=2) | -1.11690055783  0.7178535241 | -0.27 | 0.32 : -1 : 0.18 |
| C$_{10}$H$_8$ (naphthalene, N=68) | -378.683524679  457.7765564914 | -0.41 | 0.21 : -1 : 0.33 |

___

**Appendix.3 Number of operations in computation**: Both terms, i.e. the two and three center integrals in Eq.8 breaks up $K^2$ (more exactly ($^K_2$)=K(K-1)/2) cross products for all possible combinations of basis function pairs ($G_k$, $G_{k'}$), but in case of Coulomb energy, the magnitude of the number of cross products, i.e. the four center integrals in Eq.9 is $K^4$ (more exactly ($^K_2$)($^K_2$)), which can be decreased via symmetry since $\mathbf{r}_1$ and $\mathbf{r}_2$ are equivalent, etc., but the $K^4$ magnitude is huge: In case of a small system, like water dimer (H$_2$O)$_2$, there are K ~100 contracted Gaussians (using 6-311++G(3d,2p) basis set), so about $K^4$= 10$^8$ cases have to be evaluated. After calculating these cross products in Eqs.8-9 with basis set {$G_k$}, the stationer LCAO coefficients have to be found (by an HF-SCF procedure or the procedure described here) along with the multiplications and sums in Eqs.3-5 during the iteration, to obtain finally the desired energy (L) in Eq.3 to complete Eq.1. In view of computation, the origin of the difficulty with cross products can be symbolized as follows: Consider the expansion of ($\Sigma_1^K a_k$)($\Sigma_1^K b_{k'}$)= $a_1b_1$ + $a_1b_2$ + … $a_Kb_K$, the left hand side has K-1 additions in both sums plus 1 multiplication, all together 2K-1 operations, while the right hand side has $K^2$ multiplications plus $K^2$-1 additions, all together 2$K^2$-1 operations, finally about 2K vs. 2$K^2$, and similarly, for products of four sums: 4K vs. 4$K^4$. The cardinality value $K^2$ of cross products in Eq.8 is feasible, but the large $K^4$ value counting for Eq.9 is a reason to look for good approximations to make a shortcut: Numerical integration in certain circumstances [6] or approximate formulas (Appendix.4) work.

**Appendix.4 Approximate Coulomb energies**: We mention the important point in standard HF-SCF that, in case of GTO functions the terms in Eqs.8-9 can be evaluated analytically (using that for example, products of GTO is LC of GTO's which is not true for STO, etc.). In HF-SCF [1-3], the emblematic

$$V_{ee}\approx J-K= \int(j(1,2)-k(1,2))r_{12}^{-1}d\mathbf{r}_1d\mathbf{r}_2 \qquad (14)$$

approximation is used, see its particular case in Eq.1 and its derivative with $c_{ik}$ in Eq.4. The J and K integrals contain the cross terms in Eq.9, which can be evaluated analytically and before the iteration. (Eq.14 is only an approximation, so it needs "correlation energy (≈1 %)" calculation, that is, correcting the not-enough single Slater determinant (S$_0$) approximation, however, that is another question); the expression in Eq.14 has been a milestone equation in calculating Coulomb interactions on the quantum level. These famous J and K are called "Coulomb-J-integral" and "exchange-K-integral", resp.

Another (milestone) approximation (suffering also from the necessity of "exchange and correlation energy (≈1 %)" calculation) in density functional theory (DFT) [2-3] is

$$V_{ee}\approx (1/2)\int \rho(1)\rho(2)\, r_{12}^{-1}\, d\mathbf{r}_1 d\mathbf{r}_2 . \qquad (15)$$

Its derivatives ($\partial/\partial c_{ik}$)$\int \rho(1)\rho(2)r_{12}^{-1}\,d\mathbf{r}_1 d\mathbf{r}_2$= 8$\int G_k(1)$ i(1) $\rho(2)r_{12}^{-1}\,d\mathbf{r}_1 d\mathbf{r}_2$ modifies Eq.4 accordingly.

Approximations in Eqs.14-15 also suffer from the above mentioned "$K^4$ operations" problem in their analytical integration, but these analytical evaluations in Eqs.8-9 are necessary only once before the iteration starts. In ref.[4] the approximation with approximate ground state one-electron density ($\rho_0\approx \rho(1)$)

$$V_{ee}\approx C_J[\int\rho^{6/5}d\mathbf{r}_1]^{5/3} + \Sigma_{j=1,…n}C_{xj}[\int\rho^{[1+1/(3j)]}d\mathbf{r}_1]^j \qquad (16)$$

is reviewed, wherein the 1$^{st}$ term accounts for main value, and the rest is correction, even the correlation energy can be included. If Eq.16 is used for Eqs.1-5, the price is that: 1., the integrals can be evaluated numerically only (for both, GTO and STO) e.g. with the scheme used in ref.[6], 2., integrals must be evaluated in every iteration step with the improved {$c_{ik}$}U{$\lambda_{ij}$}, along using ($\partial/\partial c_{ik}$)[$\int\rho^{6/5}d\mathbf{r}_1$]$^{5/3}$~ [$\int G_k(1)i(1)\rho(1)^{1/5}d\mathbf{r}_1$]$^{2/3}$ continuing with numerical integration in the right, but another choice is a numerical derivation for $\partial/\partial c_{ik}$ also, since many similar integrands are needed anyway. The benefit from approximation in Eq.16 is that it belongs to the numerical $V_{ee}\approx \int f(\rho(1))d\mathbf{r}_1$ forms, and importantly, in fact the above mentioned "K operations" vs. analytical "$K^4$ operations" $\int(...)\,r_{12}^{-1}d\mathbf{r}_1d\mathbf{r}_2$ in Eqs.14-15, so for larger molecules Eq.16 may be useful, but a less studied area than the widely used and tested Eqs.14-15.

A more powerful approximation than Eq.16 (in which the dimensionality is $\rho^{(6/5)(5/3)}= \rho^2$) is based on Padé approximation (see Fig.1 below): For example, the two different versions are

$$V_{ee}\approx V_{ee}^{Padé-1}\equiv \int [ (\Sigma_i a_i\rho^i)/(1+ \Sigma_j b_j\rho^j) ]d\mathbf{r}_1 \quad \text{or} \quad V_{ee}\approx V_{ee}^{Padé-2}\equiv (\Sigma_i a_i\int\rho^i\, d\mathbf{r}_1)/(1+ \Sigma_j b_j\int\rho^j\, d\mathbf{r}_1). \quad (17)$$

In the right equation in Eq.17 the $\int\rho d\mathbf{r}_1=N$ is consistent with the normalization and analytical evaluation is possible with GTO, but numerical integration is necessary for STO. The left equation in Eq.17 can be evaluated numerically only for both, STO and GTO. For the parameters (4-6 terms are enough in sums) in Eq.17, a least square fit is adequate as $Y= \Sigma_n [(V(n) - V_{ee}^{Padé-1\ or\ 2}(n))w]^2$ with weight $w=1$ or $1/V(n)$, where n is e.g. about 100 small equilibrium molecules from the called G2 or G3 set [4, 7] for which $\rho= \rho_0^{HF-SCF}$ in Eq.2 is calculated via e.g. HF-SFC/basis by the "Gaussian program package". For V(n), the right hand side of Eq.14 can be used what "Gaussian program package" also calculates analytically, or alternatively the analytically evaluated Eq.15. The basis set used can be STO-3G or 6-31G** or else, its quality is not important, but it should contain only a large number (K) GTO functions yielding – hopefully - universal parameters for Eq.17. The derivation of equations, $\partial Y/\partial a_i= \partial Y/\partial b_i= 0$ for both versions in Eq.17 is straightforward. To solve this latter equation system for the left equation in Eq.17 one can use e.g. the NR method used in this work, however, the right equation in Eq.17 can be linearized for its parameters by modifying the equation for Y as follows: If $V(n)\approx (\Sigma_i a_i\int\rho^i)/(1+ \Sigma_j b_j\int\rho^j)$, then $V(n) (1+ \Sigma_j b_j\int\rho^j) - (\Sigma_i a_i\int\rho^i) \approx 0$, so let us minimize the square of the differences as $Y^{mod}= \Sigma_n[V(n) + V(n)\Sigma_j b_j\int\rho^j - \Sigma_i a_i\int\rho^i]^2$, for which $\partial Y^{mod}/\partial a_i= \partial Y^{mod}/\partial b_i= 0$ is a linear equation system for parameters $a_i$ and $b_j$. Finally, as an extension of Eq.17 for different densities (actually, Eq.9 is this kind, in fact) or between ground and excited determinants or densities [7-8], the $\int S_c^*S_d r_{12}^{-1}\Pi_{i=1}^N ds_i d\mathbf{r}_i$ or $\int\rho_c(1)\rho_d(2)r_{12}^{-1}d\mathbf{r}_1 d\mathbf{r}_2$ with c, d= 0,1,2… can be approximated with two dimensional version of Eq.17, that is, the set $\{\rho, \rho^2, \rho^3,…\}$ for LC of powers is replaced by the set $\{\rho_c, \rho_d, \rho_c^2, \rho_c\rho_d, \rho_d^2, \rho_c^3,…\}$.

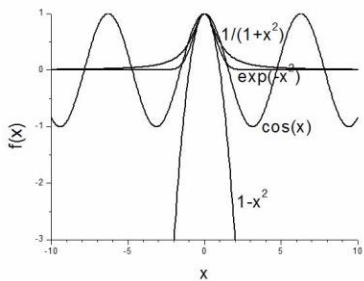

**FIGURE 1.** Schematic comparison of different (immediate, i.e. weak) fits to the vicinity of a Gaussian maximum $(\exp(-x^2))$ with Padé- $((1+x^2)^{-1})$, Fourier- $(\cos(x))$ and polynomial $(1-x^2)$ approximations to represent the unique property of a simple Padé function to recover global (asymptotic) properties in certain cases, useful when Gaussian functions are used, e.g. in describing one-electron densities.

## ACKNOWLEDGMENTS

Financial and emotional support for this research from OTKA-K- 2015-115733 and 2016-119358 are kindly acknowledged. The subject has been presented in ICNAAM 2018_39, Greece, Rhodes.